\documentstyle[preprint,aps]{revtex}
\begin{document}
\title{COSMOLOGICAL CONSTANT AND GRAVITATIONAL REPULSION EFFECT:
1. Homogeneous models with radiation}

\author{Nguyen Hong Chuong$^{\star \dag}$}
\address{Department of Physics, Syracuse University, Syracuse, NY 13244-1130,
USA}
\author{Nguyen Van Hoang}
\address{Centre for Cosmic Physics and Remote Sensing ,
20000 Hochiminh City, Vietnam}
\maketitle

\begin{abstract}
Within the framework of the minimum quadratic Poincare gauge theory of
gravity in the Riemann-Cartan spacetime we
study the influence of gravitational vacuum energy density (a cosmological
constant) on the dynamics of various gravitating systems. It is shown that
the inclusion of the cosmological term can lead to gravitational
repulsion. For some simple cases of spatially homogeneous
cosmological models with radiation we obtain non-singular solutions
in form of elementary functions and elliptic integrals.

\noindent PACS number(s): 04.50.+h, 98.80.+k
\end{abstract}

\section{Introduction}
In the Poincare gauge theory of gravity (PGT) the gravity is described
by the field of tetrads $\; h^i{}_{\mu} \;$ and the Ricci rotation
coefficients $\; A^{ik}{}_{\nu} \;$ [1-4]. Due to the dynamical independence
of these gauge fields the spacetime has both the curvature $\; F^{ik}{}_
{\mu \nu} \;$ and the torsion $\; S^i{}_{\mu \nu}.\;$
Usually, one connects the torsion with the spin angular momentum of matter.
In the simplest PGT - the Einstein-Cartan theory (ECT) [1-2] - by virtue of the
algebraic dependence between the spin angular momentum and the torsion,
 we recover Greneral Relativity (GR) for the spinless gravitating systems.

In the PGT, the choice of gravitational Lagrangian is very important. The
principle of local gauge invariance itself gives only the form of the gauge
fields and their strengths but does not allow us to determine the explicit
form of the Lagrangian for the gauge field. As was shown in [4,5] the simplest
generalization of the ECT, which satisfies the restrictions following from
the simultaneous consideration of the quantization problem (theory without
"ghosts" and "tachyons") [6], the Birkhoff's theorem [7], and the problem
of avoiding the metric singularity in the homogeneous isotropic
cosmological models [8, 9], is the minimum quadratic gauge theory of
gravity (MQGT) with the gravitational Lagrangian:
$$
L_g \; = \; h({\it f}_0F + {\alpha}F^2), \quad F_{\mu \nu} \;
= \; F^{\lambda}{}_{\mu \lambda \nu}, \quad F \; = \; F^{\mu}{}_{\mu},
\eqno(1.1)
$$
\noindent where $\; h \> = \> {\it det}(h^i{}_{\mu}), \; F \;$ is the
scalar curvature of the Riemann-Cartan spacetime, $\; {\it f}_0 \> =
\> (16{\pi}G)^{-1},\;$ (G is the Newton's gravitational constant),
and $\; \alpha \> < \> 0 \;$ is a dimensionless coefficient.

In [5] the basic field equations of MQGT were derived and in [10-14]
some of their physical consequences were investigated. In this paper
we will study the influence of the cosmological
term on the dynamics of some spatially homogeneous gravitating
systems. This work is a direct
continuation of [10] and we will use all notations in [10], unless
otherwise stated.
\section{Cosmological term and gravitational repulsion effect.}
According to the modern viewpoint, the cosmological term is connected
with the energy density of the gravitating vacuum. Taking into account
the cosmological term $\; \chi \;$ in the Lagrangian (1.1), two field
equations of MQGT are obtained by independent variation of (1.1) with
respect to $\; h^i{}_{\mu} \;$ and $\; A^{ik}{}_{\nu}$:
$$
2({\it f}_0 + 2{\alpha}F)F^{\mu}{}_i - ({\it f}_0 + {\alpha}F)Fh^{\mu}
{}_i + {\chi}h^{\mu}{}_i \; = \; - {\rm t}^{\mu}{}_i, \eqno(2.1)
$$
$$
2{\nabla}_{\nu}{\Big[}({\it f}_0 + 2 {\alpha}F) h_{[i}{}^{\mu} h_{k]}
{}^{\nu}{\Big]} \; = \; - J_{ik}{}^{\mu}. \eqno(2.2)
$$
In this work we restrict ourselves to consider ones the so-called
conformaly-invariant
gravitating systems of spinless matter with trace-free
energy-momentum tensor:
$$
J_{ik}{}^{\nu} = 0 \quad and  \quad {\rm t}^{\mu}_{\mu}
= 0. \eqno(2.3)
$$
It then follows from (2.1) that $\; F \> = \> {2\chi \over {{
\it f}_0}} \> = \> {\it const}, \;$ and consequently, from (2.2) we find
$$
S^{\lambda}{}_{\mu \nu} \; = \; 0. \eqno(2.4)
$$
After some simple manipulation from (2.1) we obtain the MQGT equation for
the considered systems in the form of the Einstein's equation of GR with
the effective energy-momentum tensor as:
$$
G^{\lambda \mu}(\{ \}) \; = \; - {1 \over {2{\it f}_0}}\>{\Big(}\>
 {\rm t}^{\lambda \mu}_{eff} + \chi g^{\lambda \mu}\>{\Big)} \; = \; -
{1 \over {2{\it f}_0}}\>{\bigg(}\>{{\rm t}^{\lambda \mu} \over {1
- 4\beta\chi}} + \chi g^{\lambda \mu}\>{\bigg)},\eqno(2.5)
$$
\noindent where $\; G^{\lambda \mu}(\{ \}) \;$ is the ordinary Einstein
tensor, $\; \beta \> = \> - {{\alpha} \over {{\it f}_0^2}} \> > \> 0, \;$
 and $\; \chi \not= {1 \over {4\beta}}. \;$ It is obvious that
if $\; \chi = 0 \;$ we recover ordinary Einstein's equation (cf.
refs [5,13]).  If $\; \chi <
{1 \over {4\beta}} \;$ the solutions of eq. (2.5) cannot provide any
qualitatively new properties in comparison with the corresponding
solutions of GR.

However, under the condition
$$
 \chi \; > \;  {1 \over {4\beta}} \eqno(2.6)
$$
\noindent
the effective energy-momentum tensor $\; {\rm t}^{\lambda \mu}_{eff} \;$
in (2.5) is negative, that leads to the gravitational repulsion effect. This
 effect is induced by a large cosmological constant (high gravitating
vacuum energy density) and by the square
curvature term in gravitational Lagrangian (1.1), although the torsion
is vanishing.
Due to this repulsion effect, the Penrose-Hawking energy conditions [15]
might be broken and we can find
a series of non-singular solutions, which are of certaint interest for
astrophysical and cosmological applications. Note that in the very
early stage of the cosmological evolution the vacuum energy density
$\; \chi \;$  was , in fact, very high.

In the following sections on the basis of (2.5) we will study the
gravitational repulsion effect in some simple spatialy homogeneous
cosmological models with ultrarelativistic perfect fluid (isotropic
radiation)
with the following energy-momentum tensor and equation of state:
$$
{\rm t}^{\lambda \mu} \; = \; ({\rho} + {\rm p}) {\rm u}^{\lambda}
{\rm u}^{\mu} - {\rm p}{\rm g}^{\lambda \mu} \eqno(2.7)
$$
$$
{\rm p} \; = \; {{\rho} \over 3} \eqno(2.8)
$$
\section{Regular Friedmann-Robertson-Walker models with
radiation.}
We consider the homogeneous isotropic cosmological models with the
Friedmann-Robertson-Walker metric:
$$
{\rm g}_{\mu \nu} \; = \; {\it diag} {\bigg(} \; 1, \; - {{R^2({\it t})}
\over {1 - kr^2}}, \; - R^2({\it t}) r^2, \; - R^2({\it t}) r^2 sin^2
{\theta} \;{\bigg)}, \eqno(3.1)
$$
\noindent
where $\; R({\it t}) \;$ is scale factor, and $\; k \> = \> - 1, \> 0,
\> + 1 \;$ for open, flat and closed models correspondingly. Then, the
00 - component of (2.5) takes the form:
$$
{k \over {R^2}} \> + {{\dot R}^2 \over {R^2}} \; = \; {1 \over {6
{\it f}_0}}{\bigg(} \> {{\rho} \over {1 - 4\beta\chi}} + \chi \> {\bigg)},
\eqno(3.2)
$$
\noindent
where a dot denotes differentiation with respect to time $\> {\it t}. \;$
The ``energy-momentum conservation law" has the standard form
$$
{\dot {\rho}} \> + \> 3 ( {\rho} + {\rm p} ) {{\dot R} \over R}
\; = \; 0. \eqno(3.3)
$$
It follows from Eqs. (3.3) and (2.8) that
$$
{\rho}\> R^4 \; = \; C_0^2 \; = \; {\it const} \; > 0. \eqno(3.4)
$$
\noindent
Subtituting (3.4) into (3.2) we obtain after simple manipulations the
differential equation for scale factor R({\it t}):
$$
{\dot R}^2 \; = \; {\chi \over {6{\it f}_0}} \> R^2 \> - \> k \>
- \> {C_0^2 \over {6{\it f}_0(4\beta\chi -1)R^2}}. \eqno(3.5)
$$
In the case considered, when $\; \chi \> > \> {1 \over {4\beta}} \;$
it is easy to find the following regular solutions
$$
R({\it t}) \; = \; {\bigg\{} \> {{3k{\it f}_0} \over {\chi}} \> + \>
{\bigg(} \> {{9k^2{\it f}_0^2} \over {{\chi}^2}} \> + \> {{C_0^2}
\over {{\chi}(4\beta\chi - 1)}}{\bigg)}^{1 \over 2} {\cosh}{\bigg[}{\bigg(}
{{2\chi} \over {3{\it f}_0}} \> {\bigg)}^{1 \over 2}({\it t - t_0}){\bigg]}
{\bigg\}}^{1 \over 2}. \eqno(3.6)
$$
\noindent
In fact, the scale factor (3.6) has the positive minimum value
$$
R_{min} \; = \; {{3k{\it f}_0} \over {\chi}} \> + \>
{\bigg(} \> {{9k^2{\it f}_0^2} \over {{\chi}^2}} \> + \> {{C_0^2}
\over {{\chi}(4\beta\chi - 1)}}{\bigg)}^{1 \over 2} \; > 0, \eqno(3.7)
$$
\noindent
not only for closed and flat models (k = 1, 0), but also for open
models, when k = - 1 by virtue $\; 4\beta\chi - 1 \>> \> 0.$ Consequently,
the energy density is limited $\; \rho \> \leq \> {\rho}_{max} \> = \>
C_0^2/R^4_{min}.$

On the basis of the obtained solution we can construct
the regular
homogeneous isotropic (inflationary) cosmological models and the regular
models of superdense isotropic radiation systems in the framework of
MQGT (cf. refs. [16, 17]).
\section{Regular Bianchi type-I models with radiation}
We now study the homogeneous Bianchi type-I radiation models with the
metric given in form:
$$
{\rm g}_{\mu \nu} \; = \; {\it diag} \> {\Big(} \> 1, \> - r_1^2({\it t}),
\> - r_2^2({\it t}), - r_3^2({\it t}) \> {\Big)}. \eqno(4.1)
$$
It follows from kk-components of eq. (2.5) (k = 1, 2, 3) the following
relationships:
$$
 r_k \; = \; {\rm r} \> {\exp} {\Big(}\>{\sl s}_k\>{\int}
{{d{\it t}} \over {{\rm r}^3}}\>{\Big)}, \eqno(4.2)
$$
\noindent
where $\; {\rm r} \; = \; (\> r_1r_2r_3 \>)^{1/3}; \; {\sum}^3_{k=1} \>
{\sl s}_k \;
= \; 0. \;$ Then, after some transformations the 00-component of eq. (2.5)
and "energy-momentum conservation law" can be written as:
$$
{{\dot {\rm r}}^2 \over {{\rm r}^2}} \; = \;{1 \over {6{\it f}_0}} \>
{\bigg(}\> {\rho \over {1 - 4\beta\chi}} \> + \> \chi \> {\bigg)} \>
+ \> {{\sl l}^2 \over {{\rm r}^6}}, \eqno(4.3)
$$
$$
{\rho}{\rm r}^4 \; = \; C_I^2 \; = \; {\it const} \; > \; 0, \eqno(4.4)
$$
\noindent
where $\; {\sl l}^2 \; = \; {1 \over 6}{\sum}^3_{k=1}\> {\sl s}_k^2, \;$
is a measure of the anisotropy. By using (4.4) it is easy
to find the exact solution of (4.3) in the following analytic form
$$
{\it t} \> - \> {\it t}_0 \; = \; {\bigg(}\>{{3{\it f}_0} \over
{2\chi}}\>{\bigg)}^{1 \over 2}\>{\int}{\bigg(}\>{{\rm x} \over {P_{I}(
{\rm x})}}\>{\bigg)}^{1 \over 2}\> d{\rm x}, \eqno(4.5)
$$
$$
P_{I}({\rm x}) \; = \; {\rm x}^3 \> - \> {{C_I^2} \over {\chi (4\beta\chi
 - 1)}}\> {\rm x} \> + \> {{6{\it f}_0{\sl l}^2} \over {\chi}}, \eqno(4.6)
$$
\noindent
where $\;{\rm x} \> = \> {\rm r}^2 \> > \> 0.\;$ It is clear that the
integral (4.5) makes sense only if $\; P_I({\rm x}) \; > 0. \;$ Note that
for all $\; {\rm x} \> \geq \> {\rm x}_1 , \;$ where $\; {\rm x}_1 \;$ is
the greatest real root of the cubic polynomial $\; P_I({\rm x}), \; P_I(
{\rm x}) \> \geq \> 0 \;$ holds. If $\; {\rm x}_1 \> \leq \> 0, \; P_I
({\rm x}) \> \geq \> 0 \; $ for all $\; {\rm x} \> \geq \> 0, \;$ and
therefore, the solution (4.5) is singular because the function $\; {\rm x}
({\it t}) \> = \> 0 \;$ at a finite time $\; {\it t} \> = \> {\it t}_0.\;$
However, if $\; {\rm x}_1 \> > \> 0 \;$ the solution (4.5) is regular:
$\; {\rm x} \> \geq \> {\rm x}_1 \> > 0 \;$ for all $\; {\it t} \> \in \>
(- \infty, \> + \infty).$
Thus, in order to get non-singular solution from (4.5) we have to find
conditions for the existence of positive real roots of the cubic
polynomial $\; P_I({\rm x}).\;$
Detailed analysis shows, that under the conditions (2.6) and
$$
{\sl l}^2 \; \leq \; A_I(C_I, \beta, \chi) \; = \;  {{C_I^3} \over
{{\sqrt {\chi}}\>[\>3(4\beta\chi -1)]^{3/2}}} \eqno(4.7)
$$
\noindent
the cubic polynomial $\; P_{I}({\rm x}) \;$ posseses three real roots $\; x_1
\>
\geq \> x_2 \> > \> 0 \> > \> x_3.\;$ Then, for $\; {\rm x} \> \geq \> x_1 \;$
the solution (4.5) is regular and can be reduced to the eliptic integrals $\;
\Pi , \> {\cal F} \;$ (see Ref. [10])
$$
{\it t \> - \> t}_0 \; = \; {\bigg(}{{6{\it f}_0} \over {x_1(x_2 - x_3){\kappa}
}}{\bigg)}^{1 \over 2}{\Big[}(x_1 - x_2)\> {\Pi}{\Big(}{\phi},\> {{x_1 - x_3}
\over {x_2 - x_3}}, \> {\kappa}{\Big)} \> + \> x_2 {\cal F} \> ({\phi},
\> {\kappa})\>{\Big]} \> , \eqno(4.8)
$$
\noindent
where
$$
\phi \; = \; {\arcsin}{\bigg[}{\bigg(} \> {{(x_2 - x_3)(x - x_1)} \over
{(x_1 - x_3)(x - x_2)}} \> {\bigg)}^{1 \over 2} \> {\bigg]}; \quad
\kappa \; = \; {\bigg(} \> {{x_2(x_1 - x_3)} \over {x_1(x_2 - x_3)}} \>
{\bigg)}^{1 \over 2} \> . \eqno(4.9)
$$
The minimum value of the universe "volume" is positive $\; V_{min} \> = \>
{\rm r}^3_
{min} \> = \> x_1^{3/2}\> > 0, \;$  where
$$
x_1 \; = \; {\bigg(}\>{{4C_I^2} \over {3\chi(4\beta\chi - 1)}}\>{\bigg)}^{1
\over 2}\>{\cos}{\bigg[}\> {1 \over 3}{\arccos}{\bigg(}
9{\sqrt{3}}{\it f}_0{\sl l}^2{\sqrt{\chi}}{(4{\beta\chi} - 1)}^{3/2}
C_I^{-3}{\bigg)}{\bigg]},   \eqno(4.10)
$$
\noindent
and the maximum value of energy density $\;{\rho}\;$ is limited: $\;
 {\rho}_{max} \> = \> {C_I^2}/{x_1^2}.$

Thus, under the conditions (2.6) and (4.7) the vacuum (repulsion) effect
dominates over ordinary gravitational contraction effect and anisotropic
effect (that is sufficiently weak), so permits avoiding the singularity
in the Bianchi type-I cosmological models with radiation.
\section{Regular Bianchi type-V models with radiation}
In this section we consider the Bianchi type-V models that are a
direct generalization of open Friedmann-Robertson-Walker models.
Taking $\;(\> {\it x}_1, \> {\it x}_2, \> {\it x}_3 \>)\;$ as local
coordinates we can write the interval of Bianchi type-V models in the
diagonal form [18, 19]
$$
ds^2 \; = \; d{\it t}^2 - r_1^2({\it t})d{\it x}_1^2 - e^{2{\it x}_1}
{\Big(}\>r_2^2({\it t})d{\it x}_2^2 + r_3^2({\it t})d{\it x}_3^2 \>
{\Big)}. \eqno(5.1)
$$
For the models considered we can find the following non-vanishing
components of the Einstein's tensor
$$
G^0_0 \; = \; {1 \over 2}{\sum}^3_{k=1}\>{{{\dot r}_k^2} \over {r_k^2}}
\> - \> {9 \over 2}{{{\dot r}^2} \over {r^2}} \> + \> {3 \over {r_1^2}} \>,
$$
$$
G^0_1 \; = \; 2 \> {{{\dot r}_1} \over {r_1}} \> - \> {\bigg(} \>
{{{\dot r}_2} \over {r_2}}
\> + \> {{{\dot r}_3} \over {r_3}} \>{\bigg)} \>,
$$
$$
G^i_i \; = \; {\bigg(}{{{\dot r}_i} \over {r_i}}{\bigg)}^{.} - 3\>{\bigg(}
 {{\dot r} \over r}{\bigg)}^{.} +3\> {{\dot r} \over r}{{{\dot r}_i} \over
{r_i}} -  {1 \over 2}{\sum}^3_{k=1}
\>{{{\dot r}_k^2} \over {r_k^2}} \> - \> {9 \over 2}{{{\dot r}^2}
\over {r^2}} \> + \> {1 \over {r_1^2}} \quad {(no \; sum)}, \eqno(5.2)
$$
\noindent
where $\; r \> = \> (r_1r_2r_3)^{1/3}. \;$
It follows from $\; G^0_1 \> = \> 0 \;$ that
$$
{{{\dot r}_1} \over {r_1}} \> = \>{1\over2}{\bigg(}{{{\dot r}_2} \over {r_2}}
 \> + \> {{{\dot r}_3} \over {r_3}}{\bigg)} \> = \> {{\dot r} \over {r}},
\eqno(5.3)
$$
\noindent
and, consequently $\; r_1 \> = \> r \;$ (by rescaling).
{}From (2.5) we get $\; G^1_1 = G_2^2 = G^3_3 \;$ and  by virtue of (5.3)
we can find
$$
r_2 \; = \; r\>{\exp}{\bigg(}\>{\sl s}\> {\int}\> {{d{\it t}} \over {r^3}}\>
{\bigg)}\>; \qquad r_3 \; = \; r\>{\exp}{\bigg(}\>- {\sl s}\> {\int}\>
{{d{\it t}} \over {r^3}}\>{\bigg)} \>, \eqno(5.4)
$$
\noindent
where $\> {\sl s} \>$ is an integration constant and has the sense of a
measure of the anisotropy. Then, 00-component of (2.5) can be reduced to
$$
{{{\dot r}^2} \over {r^2}} \; = \;{1 \over {6{\it f}_0}} \>
{\bigg(}\> {\rho \over {1 - 4\beta\chi}} \> + \> \chi \> {\bigg)} \>
+ \> {{{\sl s}^2} \over {3r^6}} \> + \> {1 \over {r^2}} \>, \eqno(5.5)
$$
The "energy-momentum conservation law" has the ordinary form
$$
{\rho} \> r^4 \; = \; C_{V}^2 \; = \; {\it const} \; > \; 0. \eqno(5.6)
$$
By analogy with the foregoing section we obtain the exact solution of (5.5)
in an analytic form
$$
{\it t} \> - \> {\it t}_0 \; = \; {\bigg(}\>{{3{\it f}_0} \over
{2\chi}}\>{\bigg)}^{1 \over 2}\>{\int}{\bigg(}\>{{\rm x} \over {P_{V}(
{\rm x})}}\>{\bigg)}^{1 \over 2}\> d{\rm x} \; , \eqno(5.7)
$$
$$
P_{V}({\rm x}) \; = \; {\rm x}^3 \> + {{6{\it f}_0} \over {\chi}}\> {\rm x}
^2 \> - \> {{C_{V}^2} \over {\chi (4\beta\chi
 - 1)}}\> {\rm x} \> + \> {{2{\it f}_0{\sl s}^2} \over {\chi}}, \eqno(5.8)
$$
\noindent
where $\;{\rm x} \> = \> {\rm r}^2 \> > \> 0.\;$

By analogy with
section 4, together with condition (2.6) we can
find the similar to (4.7) condition for the existence of positive real root
of the cubic polynomial $\; P_{V}({\rm x}) \;$ as
$$
{\sl s}^2 \; \leq \; A_V(C_V, \beta, \chi) \; = \;   x_+ \> {\bigg(} \>
{{C_V^2} \over {3{\it f}_0(4\beta\chi -1)}} \> - \> x_+ \> {\bigg)}, \eqno(5.9)
$$
\noindent
where
$$
 x_+ \> = \> {\bigg(}{{4{\it f}_0^2} \over {{\chi}^2}} \> + \>
{{C_V^2} \over {3\chi(4\beta\chi - 1)}}\> {\bigg)}^{1 \over 2} \> - \>
{{2{\it f}_0} \over {\chi}}\> .\eqno(5.10)
$$
It is easy to show that $\; A_V(C_V, \beta,
 \chi) \> > \> 0, \;$ and consequently, the condition (5.9) makes sense.
Note that condition (5.9) gives stronger a restriction on the
anisotropy than (4.7).
In the case considered, the vacuum (repulsion) effect dominates over the
ordinary gravitational attraction effect, the anisotropic effect, and
the effect of negative curvature.

Thus, under the conditions (2.6) (large cosmological constant) and (5.9)
(small anisotropy) the solution (5.7) is regular and has the same form
as (4.8). Then, we can construct the non-singular Bianchi type-V radiation
cosmological models with the limiting energy density $\; {\rho} \> \leq \>
{\rho}_{max} \> < \> {\infty}, \;$ and nonzero metric functions $\; r_i(
{\it t}) \> > \> 0 \;$ at any time.
\section{Conclusions}
In this paper we have shown that the gravitating vacuum effect can be
of importance for avoiding the singularity in some homogeneous cosmological
models with the isotropic radiation. We obtain regular solutions for
Friedmann-Robertson-Walker, Bianchi type-I, and Bianchi type-V models.
These solutions have been expressed in terms of elementary functions and
 elliptic integrals.
Note that the gravitational repulsion effect can play an important role
in the case of conformally-invariant systems such as: massless
conformally-invariant scalar field, electromagnetic field... On the basis
of eq. (2.5) following [20] we can find exact non-singular solutions
for various cosmological models with radiation, scalar, and electromagnetic
fields, as well as for spherically-symmetric conformally-invariant systems.

\acknowledgements
The authors are grateful to Profs. A. V. Minkevich and F. I, Fedorov for
their guidance during their study in Minsk. N. H. Chuong would like to
thank Profs. S. Bazanski,  and
W. Kopczynski for warm hospitality in Warsaw, where the first draft
of the paper was presented. He would especially like to thank Prof.
A. Zichichi and the World Laboratory for providing the fellowship and
Prof. A. Ashtekar for kind attention during his stay in Syracuse. This work
was supported in part by research funds provided by Syracuse University.

\end{document}